\documentclass[pra,amssymb,amsmath,showpacs]{revtex4}
\usepackage{graphicx}

\newtheorem{thm}{Theorem}
\newtheorem{crl}{Corollary}
\def\qed{$\Box$}
\begin{document}

\title{Non-destructive      Orthonormal      State     Discrimination}
\author{M.  Gupta}  \email{er_manugupta@yahoo.com} \author{A.  Pathak}
\email{anirban.pathak@jiit.ac.in}   \affiliation{Jaypee  Institute  of
Information  Technology, Noida, 201  307, India}  \author{R. Srikanth}
\email{srik@rri.res.in}    \affiliation{Raman    Research   Institute,
Bangalore,    560    012,     India}    \author{P.    K.    Panigrahi}
\email{prasanta@prl.ernet.in}      \affiliation{Physical      Research
Laboratory, Navrangpura, Ahmedabad, 380 009, India}

\begin{abstract}
We   provide  explicit  quantum   circuits  for   the  non-destructive
deterministic  discrimination  of Bell  states  in  the Hilbert  space
$C^{d^{n}}$, where  $d$ is  qudit dimension. We  discuss a  method for
generalizing  this  to  non-destructive  measurements on  any  set  of
orthogonal states  distributed among $n$ parties.   From the practical
viewpoint,  we show  that such  non-destructive measurements  can help
lower quantum communication complexity under certain conditions.
\end{abstract}
 \maketitle

\section{Introduction}
Entangled states play a key role in the transmission and processing of
quantum information \cite{Niel,suter}.  Using an entanglement channel,
an  unknown state  can be  teleported \cite{bouw}  with  local unitary
operations, appropriate  measurement and classical  communication; one
can  achieve entanglement  swapping through  joint measurement  on two
entangled  pairs \cite{pan1}.  Entanglement leads  to increase  in the
capacity of  the quantum information  channel, known as  quantum dense
coding \cite{Mattle}.  The bipartite, maximally  entangled Bell states
provide the  most transparent illustration of  these aspects, although
three particle entangled states like GHZ and W states are beginning to
be employed for various purposes \cite{carvalho,hein}.

Making use  of single  qubit operations and  the C-NOT gates,  one can
produce various entangled states in a quantum network \cite{Niel}.  It
may be of interest to know the type of entangled state that is present
in a  quantum network,  at various stages  of quantum  computation and
cryptographic      operations,      without      disturbing      these
states.  Nonorthogonal states cannot  be discriminated  with certainty
\cite{wootters},  while the  discrimination of  orthogonal  states are
possible. A  large number of results  regarding distinguishing various
orthogonal     states,      have     recently     been     established
\cite{walgate,gosh1,vermani,chen}. If two copies belonging to the four
orthogonal Bell  states are  provided, local operations  and classical
communication   (LOCC)  can   be   used  to   distinguish  them   with
certainty. It is not possible  to discriminate using only LOCC, either
deterministically or probabilistically among  the four Bell states, if
only a single  copy is provided \cite{gosh1}. It  is also not possible
to  discriminate multipartite  orthogonal  states by  using LOCC  only
\cite{gosh2}. However,  any two multipartite orthogonal  states can be
unequivocally distinguished through LOCC \cite{walgate}.

A number of theoretical and experimental results already exist in this
area         of        unambiguous         state        discrimination
\cite{cola,pan2,kim}. Appropriate unitary transforms and measurements,
which  transfer the Bell  states into  disentangled basis  states, can
unambiguously     identify     all     the    four     Bell     states
\cite{pan2,kim,boschi}.   However, in the  process of  measurement the
entangled state  is destroyed.  Of  course, the above  is satisfactory
when the Bell state is not required further in the quantum network.

We consider in this work  the problem of discriminating a complete
set of orthogonal  basis states in  $C^{d^n}$-- of which  the
conventional Bell  states form  a special  case-- where  the $n$
qudits ($d$-level systems) are distributed among $n$ players.  We
present a scheme which deterministically   discriminates   between
these   states   without vandalizing them, such that these are
preserved for further use.  This article is divided as  follows.  In
Section \ref{sec:osd2}, we present circuits  for the non-destructive
Bell  state discrimination  for $n$ qubits  shared among  $n$
players,  beginning  with   the  case  of conventional Bell states.
In Section \ref{sec:diskri}, this result is generalized to construct
circuits  for Bell state discrimination among qudits.   In  Section
\ref{sec:gen},  we  point  out  the underlying mathematical
structure  that  clarifies  how  our proposed  circuits work. In
principle, this can be used to further generalize our results of
Section \ref{sec:diskri}  to discrimination  of any set  of
orthogonal states.   In  Section \ref{sec:pli},  we  examine
specific  situations where such non-destructive measurements  can be
useful in  computing and cryptography. An appendix is attached at
the end, which shows closure property of generalized Bell states,
used in the text under Hadamard operations.

\section{Bell state discrimination in $C^{2^{n}}$ Hilbert space
\label{sec:osd2}} In principle,  any set  of orthogonal states  can
be  discriminated in quantum  mechanics, but LOCC  may not  be
sufficient  if the  state is distributed  among  two  or  more
players.   Here  we  start  with  a $C^{2^{n}}$ Hilbert space. To
describe any state in this Hilbert space we need $2^{n}$ orthonormal
basis vectors.  The choice of the basis is not  unique, but one
choice of  particular importance  is the  set of maximally entangled
$n$-qubit generalization of Bell states given by:
\begin{subequations}
\label{eq:basis}
\begin{eqnarray}\label{eq:basisa}
|\psi_{x}^+\rangle & = & \frac{1}{\sqrt{2}}(|x\rangle+|\bar{x}\rangle),\\
|\psi_{x}^-\rangle & = &
\frac{1}{\sqrt{2}}(|x\rangle-|\bar{x}\rangle)
\label{eq:basisb}
\end{eqnarray}
\end{subequations}
where $x$ varies from 0  to $2^{n-1}-1$ and $\bar{x} \equiv 1^{\otimes
n} \oplus x$ in modulo 2 arithmetic. The set of complete basis vectors
(\ref{eq:basis}) reduces to Bell basis for $n=2$ and to GHZ states for
$n=3$. As  an example,  setting $n=2$ in  (\ref{eq:basis}) we  get the
usual Bell states
\begin{eqnarray}
|\psi_{00}\rangle=|\psi^{+}\rangle & = &
\frac{1}{\sqrt{2}}(|00\rangle+|11\rangle),\nonumber \\
|\psi_{01}\rangle=|\phi^{+}\rangle & = &
\frac{1}{\sqrt{2}}(|01\rangle+|10\rangle),\nonumber \\
|\psi_{10}\rangle=|\psi^{-}\rangle & = &
\frac{1}{\sqrt{2}}(|00\rangle-|11\rangle),\nonumber \\
|\psi_{11}\rangle=|\phi^{-}\rangle & = &
\frac{1}{\sqrt{2}}(|01\rangle-|10\rangle).\label{eq:bell}
\end{eqnarray}

\begin{figure}
\includegraphics[width=5in]{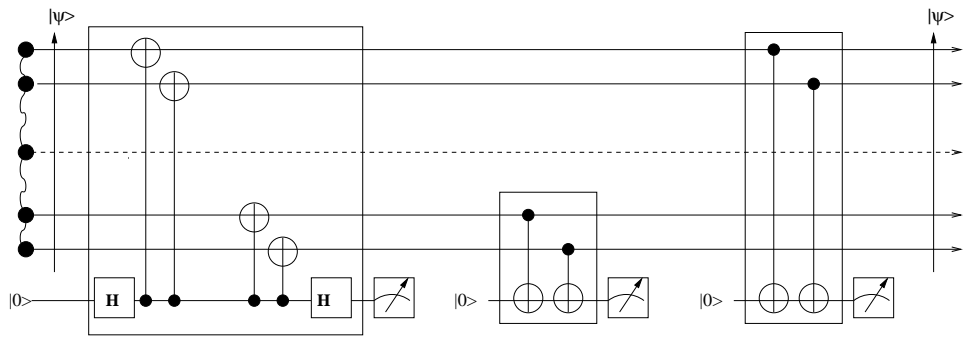}
\caption{Diagram depicting the circuit for non-destructive generalized
orthonormal  qubit Bell  state  discriminator. The  first bounded  box
depicts, in a sense  clarified in Section \ref{sec:gen}, an effective,
non-destructive measurement of $X^{\otimes n}$, which yields the phase
bit  value.   The  second  and   third  boxes  depict   an  effective,
non-destructive measurement  of $Z^{\dag}\otimes Z$,  which yields the
relative parity  between two consecutive  qubits.  To obtain  the full
relative  parity information, $n-1$  relative parity  measurements are
required.}
\label{fig:gsd}
\end{figure}

A   circuit   to   non-destructively  discriminate   the generalized
orthonormal entangled basis  states (\ref{eq:basis}) employing
ancilla is shown in Fig. \ref{fig:gsd}.   To discriminate the
members of the entangled,  orthonormal   basis  set   in
$C^{2^{n}}$,  we have  to communicate and carry out measurements on
$n$ ancillary qubits in the computational  basis.  The  first
measurement is  done on the  state $|R_{nA_1}\rangle$,  as shown in
Eq. (\ref{gen  state dis1}).   This measurement  determines  the
relative  phase  between $|x\rangle$  and $|\bar{x}\rangle$. It
will       give $0$       for
$\frac{1}{\sqrt{2}}(|x\rangle+|\bar{x}\rangle)$ and     $1$ for
$\frac{1}{\sqrt{2}}(|x\rangle-|\bar{x}\rangle)$. The next
measurements compare the parity between two  consecutive bits and
yield zero if the bits coincide  and one, otherwise.   This follows
from  Eq.  (\ref{gen state dis2}), which shows the state for the
complex of the system and the $i$th  ancilla, where  $2 \leq i \leq
n$.  Each ancilla  $A_i$ is sequentially interacted with the system
and then measured.  It can be shown (Section \ref{sec:diskri}) that
this action leaves  the states $|\psi^{\pm}_x\rangle$ undisturbed.
This means that the corresponding measurements,  $M_i$, represent
commuting observables.  In  general, $M_1$ gives  the phase bit, and
$M_i$ gives the parity  of the string comprising of the $i$th and
$i{+}1$th qubits.

In a way clarified in  Section \ref{sec:gen}, $M_1$ may be regarded as
the non-destructive equivalent of  measuring $X^{\otimes n}$ and $M_i$
($2  \le i  \le  n$) that  of measuring  $Z  \otimes Z$,  so that  the
simultaneous  measurability of any  pair of  $M_i$'s follows  from the
fact that  $[X^{\otimes n}, Z(j)\otimes  Z(k)] = 0$  and $[Z(j)\otimes
Z(k), Z(j^{\prime})\otimes Z(k^{\prime})]=0$ where $Z(j)$ is the Pauli
$Z$ operator acting on the $j$th qubit.

A  note on  notation: the  sign $Q(j\leftarrow  k)$ signifies  a C-NOT
gate, with  $k$ being  (ancilla) control index  number, and  $j$ being
(system)  target   index  number.   Conversely,   $Q(j\rightarrow  k)$
signifies a  C-NOT gate with  $j$ being (system) control  index number
and $k$ being (ancilla) target index number.
\begin{subequations}
\label{eq:aftan}
\begin{eqnarray} \label{gen state dis1}
|R_{(n\times2)A1}\rangle & = & \left[I_2^{\otimes n}\otimes H_2\right]\times
\left[\bigotimes_{j=1}^{n}Q(j\leftarrow1)\right]
\times \left[I_2^{\otimes n}\otimes H_{2}\right]
(|\Psi\rangle_{1\cdots n}\otimes|0\rangle_{A1}), \\
\label{gen state dis2} |R_{(n\times2)Ai}\rangle &
= & \left[Q([i-1]\rightarrow i)
\otimes Q\left([{i}\rightarrow{i}\right)\right]
\left(|\Psi\rangle_{1\cdots n}\otimes|0\rangle_{Ai}\right),
\end{eqnarray}
\end{subequations}
where  $2  \le  i \le  n-1$.   Therefore,  all  together we  need  $n$
measurements  on   $n$  ancillary  qubits   to  discriminate  $2^{n}$
orthonormal,     entangled     basis     states    of     the     form
(\ref{eq:basis}). Furthermore, we  require $3n-2$ applications of CNOT
gates.  The question  of quantity  of quantum  communication required,
which depends  on the topology  of the quantum  communication network,
is discussed in Section \ref{sec:pli} in detail.

\begin{figure}
\includegraphics[width=5in]{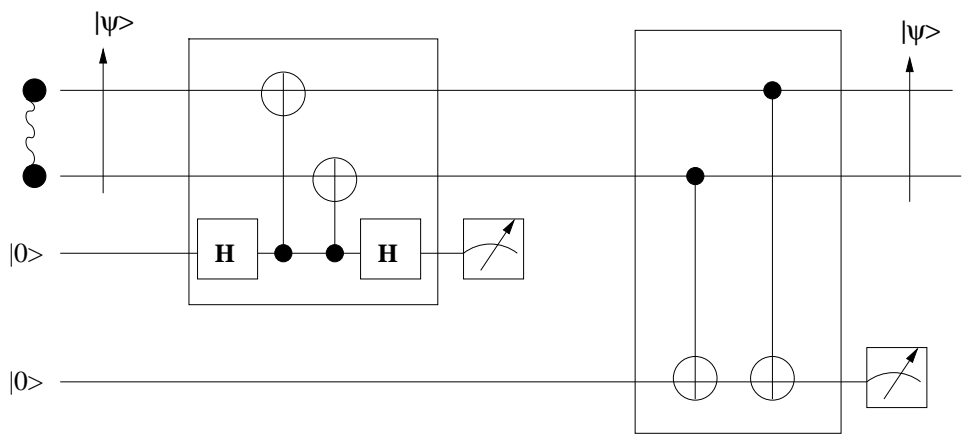}
\caption{A special  case of Fig. \ref{fig:gsd}:  the diagram
depicting the circuit for Bell state discriminator.} \label{fig:bsd}
\end{figure}

A  proof  that the  circuit  described  in  Eq. (\ref{eq:aftan}),
and depicted  in Fig. \ref{fig:gsd}  achieves the  required Bell
state discrimination  is  deferred  to  Section \ref{sec:diskri}.
Here  we simply  illustrate it  using the  specific example  of the
usual Bell states    (\ref{eq:bell}).   Since    (\ref{eq:basis})
reduces   to (\ref{eq:bell})  for $n=2$,  our generalized  circuit
reduces  to that shown  in  Fig. \ref{fig:bsd},  where  one  needs
only two  ancillary qubits, four  CNOT gates, two  measurements and
two qubits  of quantum communication.

In  Table  \ref{tab:bellsta},  we   have  shown  the  results  of  the
measurements  on both  the  ancillas when  different  Bell states  are
present  in  the  given  circuit  (Fig.  \ref{fig:bsd}).  Just  before
measurement, the states can be explicitly written as,
\begin{subequations}
\label{eq:bst}
\begin{eqnarray} \label{bst1}
|R_{(2\times2)A1}\rangle & = & \left[I_{2}\otimes I_{2}\otimes H_{2}\right]
\times[Q(1\leftarrow1)\otimes Q(2\leftarrow1)]\times
\left[I_{2}\otimes I_{2}\otimes H_{2}\right]
(|\Psi\rangle_{12}\otimes|0\rangle_{A1})\\
\label{bst2}
|R_{(2\times2)A2}\rangle & = &
\left[Q(1\rightarrow2)\otimes Q(2\rightarrow2)\right]
\left(|\Psi\rangle_{12}\otimes|0\rangle_{A2}\right).
\end{eqnarray}
\end{subequations}

\begin{table}[h]
\begin{tabular}{|c|c|c|}
\hline \hline
{\bf Bell State} & {\bf Measurement $A_1$} & {\bf Measurement $A_2$} \\
\hline
  $|\psi^+\rangle$ & $0$ & $0$ \\
  $|\psi^-\rangle$ & $1$ & $0$ \\
  $|\phi^+\rangle$ & $0$ & $1$ \\
  $|\phi^-\rangle$ & $1$ & $1$ \\ \hline
\end{tabular}
\caption{Results of outsourced measurements on two ancilla for the
Bell states \ref{eq:bell}}. \label{tab:bellsta}
\end{table}
Thus we have provided a circuit for orthonormal qubit Bell state
discrimination shared between two or more parties. These results
can be straightforwardly generalized, as shown in the following
Section.

\section{Generalized Bell state discrimination in $\mathbb{C}^{d^n}$
\label{sec:diskri}} The results of  the preceding Section can be
generalized to entangled states  of $n$  qudits.  To  this end,  we
replace  the  regular Pauli matrices   with  their   $d$-dimensional
analogs   \cite{kni96}.   We generalize $X$  and $Z$ gates; these
denoted by $X_d$  and $Z_d$, respectively, have the action:
\begin{subequations}
\label{eq:genzx}
\begin{eqnarray}
\label{eq:genzxa}
Z_d|j\rangle &\mapsto& e^{2\pi\iota j/d}|j\rangle \\
\label{eq:genzxb}
X_d|j\rangle &\mapsto& |j-1\rangle,
\end{eqnarray}
\end{subequations}
where  the  increment  in the  ket  is  in  mod $d$  arithmetic.   The
operators $X_d$  and $Z_d$ are related  by a Fourier  transform $X_d =
H_dZ_dH_d^{\dag}$,   where   $H_d$   is   the   generalized   Hadamard
transformation given by:
\begin{equation}
(H_d)_{jk}=\frac{1}{\sqrt{d}}e^{2\pi \iota j\cdot k/d}.
\label{eq:gen hadamard}
\end{equation}
Unlike the qubit case, $Z_d, X_d$ and $H_d$ are not Hermitian.

The $d$ generalized Bell states are
\begin{equation}
|\Psi_{pq}\rangle=\frac{1}{\sqrt{d}}\sum_{j}e^{2\pi\iota j p/d}
|j\rangle|j+q\rangle,\hspace{1.0cm}(0 \le p,q \le d-1)
\label{eq:genbell}
\end{equation}
which  form  an  orthogonal,  complete basis  of  maximally
entangled vectors  for the  $d^2$ dimensional  "qudit" space
\cite{ben93}. The parameter  $p$ denotes  phase  and $q$  the
generalized parity.   The states $|\Psi_{pq}\rangle$ are
$d$-dimensional analogs of Bell states (\ref{eq:bell})  in that they
are  eigenstates of  the  operator $X_d \otimes  X_d$, which is
equivalent to  the  phase observable,  whose eigenvalues are $p$ or
some function $f(p)$,  and $Z_d^{\dag} \otimes Z_d$, which is
equivalent  to the parity observable, whose eigenvalues are $q$ or
some  real-valued function $f(q)$.  Therefore, measurements
equivalent to these operators guarantee a complete characterization
of the generalized Bell states.  Furthermore, the set of generalized
Bell states remains closed under the  action $H^{\dag}_d \otimes H$
or $H_d \otimes H_d^{\dag}$ or (cf. Appendix \ref{dx:hh}).

The generalization of the CNOT that  we require is the one, whose
action we define by,
\begin{equation}
{\cal C}_X: |j\rangle|k\rangle \longmapsto |j\rangle|j-k\rangle.
\label{eq:c-}
\end{equation}
The reason for this choice  is clarified in Section \ref{sec:gen}.  We
use  the  following notation:  the  sign  ${\cal C}_X(j\leftarrow  k$)
signifies a C-SUM gate with  $k$ being (ancilla) control index number,
and $j$  being (system) target index  number; ${\cal C}_X(j\rightarrow
k$) signifies a  C-SUM gate with the control-target  order reversed. A
similar terminology extends to the two-qudit gate ${\cal C}_X^{\dag}$,
whose  action  is  given  by  either  $|j\rangle|k\rangle  \longmapsto
|j\rangle|k-j\rangle$      or      $|j\rangle|k\rangle     \longmapsto
|j-k\rangle|j\rangle$, depending  on whether the system  or ancilla is
the control register.

A direct generalization to $d$-dimension of Eq. (\ref{eq:bst}) is
\begin{subequations}
\label{eq:cmin}
\begin{eqnarray}
\label{eq:cmina}
|R_{(2\times d)A1}\rangle & = & [I_{d}\otimes I_{d}\otimes
H_{d}]\times \left[{\cal C}_X(1\leftarrow1)
{\cal C}_X(2\leftarrow1)
\right]\times
  [I_{d}\otimes I_{d}\otimes H_{d}^{\dagger}](|\Psi\rangle_{12}
\otimes|0\rangle_{A1}).\\
\label{eq:cminb}
|R_{(2\times d)A2}\rangle & = & \left[{\cal C}_X
(1\rightarrow2){\cal C}_X^{\dag}(2\rightarrow2)\right]
(|\Psi\rangle_{12}\otimes|0\rangle_{A2}).
\end{eqnarray}
\end{subequations}
We   will   denote   the   observables   corresponding   to   circuits
(\ref{eq:cmina})   and   (\ref{eq:cminb})    as   $M_1$   and   $M_2$,
respectively. $M_1$  will yield the  `phase value' $p$, and  $M_2$ the
generalized parity, $q$.  In a way clarified in Section \ref{sec:gen},
$M_1$ and  $M_2$ correspond,  respectively, to the  unitary operations
$X\otimes  X$  and  $Z^{\dag}\otimes  Z$,  so  that  the  simultaneous
measurability of $M_1$ and $M_2$ can  be shown as a consequence of the
fact that $[X\otimes X,  Z^{\dag}\otimes Z]=0$. More directly, we will
show  that  both  measurements  leave  the  state  $|\Psi_{pq}\rangle$
undisturbed.

Let  us now  consider  the more  general  system of  $n$ qudits.
The elements of  the $d^n$  dimensional vector space  over the
modulo $d$ field is given  by the set $V_d^{\times n} \equiv  \{{\bf
x}_j = (x_1, x_2, \cdots, x_n)\}$. Consider the equivalence relation
given by ${\bf x}_j  \equiv {\bf  x}_k$ if  and only  if ${\bf
x}_j-{\bf y}_k$  is a uniform vector, i.e., one of the form
$(r,r,r,\cdots,r)$, where $r \in \{0, 1,  2, \cdots, d-1\}$.  There
are  $d^{n-1}$ equivalence classes, uniquely  labeled by  the
coordinates $(q_1,q_2,\cdots,q_{n-1})  \in V_d^{\times (n-1)}$.   A
complete, maximally entangled  Bell basis for the Hilbert space
$\mathbb{C}^{d^n}$ can be given by:
\begin{equation}
|\Psi_{pq_1q_2\cdots  q_{n-1}}\rangle  =  \sum_{j=0}^{d-1}e^{2\pi\iota
j\cdot p/d} |j,q_1+j,q_2+j,\cdots,q_{n-1}+j\rangle.
\end{equation}
We   call   them  Bell   states   in   the   sense  that   any   state
$|\Psi_{pq_1q_2\cdots    q_{n-1}}\rangle$   is   an    eigenstate   of
$X_d^{\otimes n}$  and $Z_d(j)\otimes  Z_d^{\dag}(j+1)$ ($1 \le  j \le
(n-1)$), which, in a way clarified in Section \ref{sec:gen},
correspond to observables with  eigenvalues $p$
and $q_{j+1}-q_j$ respectively, the latter being called the {\em
relative parity}.

\begin{figure}
\includegraphics[width=6in]{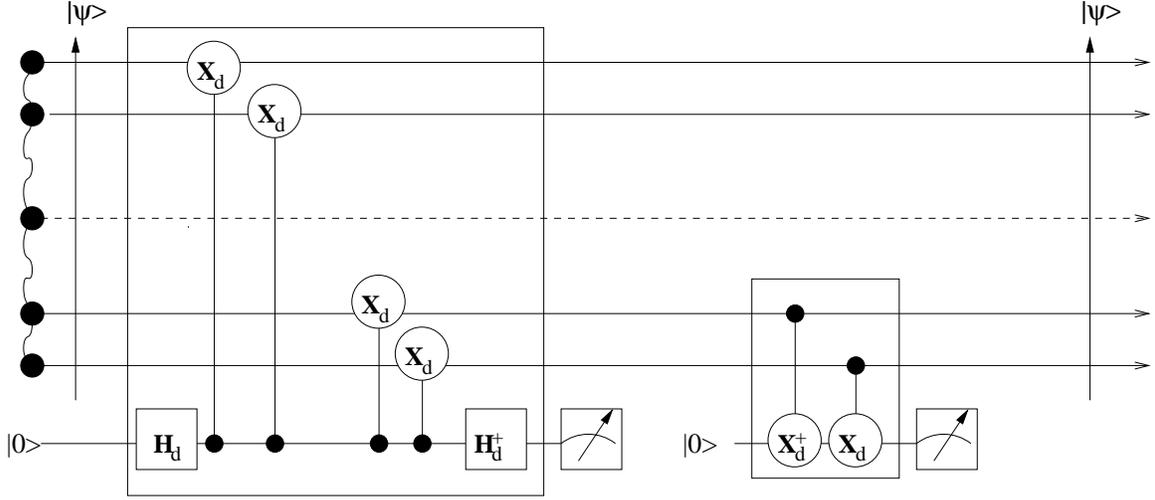}
\caption{Diagram depicting the circuit for non-destructive generalized
orthonormal qudit Bell state discriminator. The first bounded box depicts
the outsourced measurement of an observable that is compatible
with $X_d^{\otimes n}$, which
for the generalized Bell states yields the global phase value $p$.
The second box
depicts the outsourced measurement of an observable compatible
with $Z_d^{\dag}\otimes Z_d$,
which yields the relative parity between two consecutive
qudits. To obtain the full relative parity information, $n-1$ such
relative parity measurements are needed.}
\label{fig:dgsd}
\end{figure}

A  generalization  of  Eq.   (\ref{eq:cmin})  to  $n$  qudits  is
Eq. (\ref{eq:aftann}),  which   describes  a  circuit   to  measure
phase information $p$  and generalized parity  information $q_1,
q_2,\cdots, q_{n-1}$  of   such  states.   The  circuit  is depicted
in  Fig. \ref{fig:dgsd}.
The required ancilla
are $n$ qudits.
The corresponding equation is obtained by generalizing Eqs. (\ref{eq:aftan}).
\begin{subequations}
\label{eq:aftann}
\begin{eqnarray} \label{aftanna}
|R_{(n\times d)A1}\rangle & = & \left[I_d^{\otimes n}\otimes
H_d^{\dag}\right]\times
\left[\Pi_{j=1}^{n}{\cal C}_X(j\leftarrow1)\right]
\times \left[I_d^{\otimes n}\otimes H_{d}\right]
(|\Psi\rangle_{1\cdots n}\otimes|0\rangle_{A1}), \\
|R_{(n\times d)Ai}\rangle & = &
\left[{\cal C}_{X_d}([i-1]\rightarrow i)
{\cal C}_{X}^{\dag}(i\rightarrow i) \right]
\left(|\Psi\rangle_{1\cdots n}\otimes|0\rangle_{Ai}\right).\label{aftannb}
\end{eqnarray}
\end{subequations}
We  will  denote the  measurements  realized  by  these circuits,  via
ancilla $A_i$,  by $M_i$ ($1 \le i  \le n$).  To see  that the $M_i$'s
are compatible, and that  therefore their actions are non-destructive,
it turns out  to be sufficient to note  that $[X_d^{\otimes n}, Z_d(j)
\otimes  Z_d^{\dag}(k)]   =  0$   ($j  \ne  k$)   and  $[Z_d(j)\otimes
Z_d^{\dag}(k), Z_d(j^{\prime})\otimes Z_d^{\dag}(k^{\prime})] = 0$ ($j
\ne k$,  $j^{\prime} \ne k^{\prime}$),  which indeed follows  from the
fact the states $|\Psi_{pq_1q_2\cdots q_{n-1}}\rangle$ are eigenstates
of  $X_d^{\otimes n}$  and  $Z_d^{\dag}(j) \otimes  Z_d(k)$.  We  show
below  explicitly  that   the  $M_i$'s  measure  $|\Psi_{pq_1q_2\cdots
q_{n-1}}\rangle$ non-destructively.

To see this, we note
that the action of the first two (boxed) operations in Eq. \ref{aftanna}
on a state $|\Psi_{pq_1q_2\cdots q_{n-1}}\rangle|k\rangle$ is
\begin{eqnarray}
\label{eq:fasverk}
|\Psi_{pq_1,q_2,\cdots,q_{n-1}}\rangle|k\rangle &=&
\left[\sum_{j=0}^{d-1}e^{2\pi\iota j\cdot p/d}|j,q_1+j,q_2+j,\cdots,q_n+j
\rangle\right]|k\rangle \nonumber \\
&\longrightarrow& \left[\sum_{j=0}^{d-1}e^{2\pi\iota j\cdot p/d}
|j,q_1+j-k,q_2+j-k,\cdots,q_n+j-k\rangle\right]|k\rangle \nonumber\\
&=&
\left[\sum_{j^{\prime}=0}^{d-1}e^{2\pi\iota j^{\prime}\cdot p/d}
|j^{\prime},q_1+j^{\prime},q_2+j^{\prime},\cdots,q_n+j^{\prime}\rangle
\right]|k\rangle \nonumber \\
&=& e^{2\pi\iota k\cdot p/d}|\Psi_{pq_1,q_2,\cdots,q_{n-1}}\rangle|k\rangle,
\end{eqnarray}
from which it follows that full effect of the operation
described in Eq. (\ref{aftanna}) produces the state:
\begin{eqnarray}
|\Psi_{pq_1,q_2,\cdots,q_{n-1}}\rangle H_d |k\rangle &=&
|\Psi_{pq_1,q_2,\cdots,q_{n-1}}\rangle\left(\frac{1}{\sqrt{d}}\sum_{j=0}^{d-1}
|j\rangle\right) \nonumber \\
&\longrightarrow&
|\Psi_{pq_1,q_2,\cdots,q_{n-1}}\rangle\left(\frac{1}{\sqrt{d}}\sum_{j=0}^{d-1}
e^{2\pi\iota p\cdot j/d}|j\rangle\right) \nonumber \\
&\longrightarrow& |\Psi_{pq_1,q_2,\cdots,q_{n-1}}\rangle|p\rangle.
\end{eqnarray}
This yields the phase bit upon the ancilla being measured.

It is easily seen that the action (\ref{aftannb}) non-destructively
extracts the relative parity information. For,
\begin{eqnarray}
\lefteqn{\left[{\cal C}_{X_d}([i-1]\rightarrow i)
{\cal C}_{X_d}^{\dag}(i\rightarrow i)\right]
|\Psi_{pq_1,q_2,\cdots,q_{n-1}}\rangle|0\rangle_i }
\hspace{0.5cm}\nonumber \\
&=& {\cal C}_{X_d}([i-1]\rightarrow i)
\sum_{j=0}^{d-1}e^{2\pi\iota j\cdot p/d}
|j,q_1+j,q_2+j,\cdots,q_{n-1}+j\rangle|q_{i+1}+j\rangle_i\nonumber \\
&=& \sum_{j=0}^{d-1}e^{2\pi\iota j\cdot p}
|j,q_1+j,q_2+j,\cdots,q_{n-1}+j\rangle|q_{i+1}-q_i\rangle_i\nonumber \\
&=& |\Psi_{pq_1,q_2,\cdots,q_{n-1}}\rangle|q_{i+1}-q_i\rangle_i.
\end{eqnarray}
The   operation   $\left[{\cal   C}_{X_d}([i-1]\rightarrow  i)   {\cal
C}_{X_d}^{\dag}(i\rightarrow  i)\right]$ serves  to entangle  and then
disentangle  the input  Bell  state  and the  ancilla,  such that  the
relative parity of the two concerned qudits can be read off the latter
in the computational  basis. This also proves that  the circuits given
in Eqs.  (\ref{eq:aftan}), (\ref{eq:bst}) and  (\ref{eq:cmin}) perform
non-destructive  Bell  state discrimination  in  dimensions $2^n$,  $2
\times 2$  and $d  \times d$, respectively,  for they are  all special
cases of the circuit described in Eq. (\ref{eq:aftann}).

\begin{figure}
\includegraphics[width=5in]{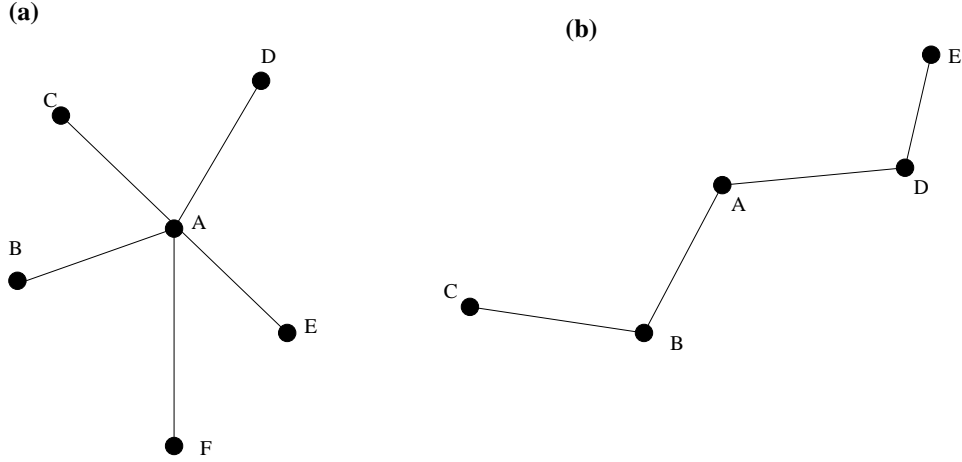}
\caption{Two possible configurations of the quantum communication network:
(a) In a star topology, a set of `relative parity' measurements could
be along each edge; (b) in the linear configuration, the strategy of
observing consecutive qubits or qudits, as given in
Eqs. (\ref{gen state dis2}) or (\ref{eq:cminb}), can be used.}
\label{fig:topo}
\end{figure}

Note that although the circuit  for qubits in Fig. \ref{fig:gsd} and
for qudits  in Fig.\ref{fig:dgsd} use  relative parity measurements
on consecutive pairs of qudits, they need not do so. Given any set
of $n-1$ relative parity values $q_j-q_k$ that suffice to fully
determine the  $q_j$'s in  a state  $|\Psi_{pq_1q_2\cdots
q_{n-1}}\rangle$, our non-destructive measurements are such that the
generalized Bell states are eigenstates  of such operators, and
hence form a  complete set of compatible observables.   In Section
\ref{sec:gen}, we  show that such relative  parity measurements
correspond  to an  observable compatible with $Z^{\dag}_d(j)\otimes
Z_d(k)$ (in  the $d=2$ case, the observable is identical with  $Z(j)
\otimes Z(k)$). Depending on  the topology of the quantum
communication network  available, the choice  of relative parity
measurements  can vary.   For  example,  if the  communication
network has a star topology,  as in Fig. \ref{fig:topo}(a), then the
set  of observables can  correspond to  $Z^{\dag}_d(1)\otimes
Z_d(j)$, where 1  is the  hub index (marked  $A$ in  the figure),
and  $j$ runs through  the  remaining vertices.   Since  any  of the
operators $X_d^{\otimes  n}$  and   $Z^{\dag}_d(j)\otimes  Z_d(k)$
commute,  by corollary    \ref{crl:utsrc}    (in    Section
\ref{sec:gen})    the non-destructive versions  of measurements
compatible with  them can be simultaneously determined.

\section{General circuits for non-destructive orthonormal
state discrimination \label{sec:gen}}

In  this Section,  we will  examine the  basic  mathematical structure
underlying our  circuits. In so  doing, we will  be able to  adapt the
ideas of the  preceding Sections to the case  of any orthonormal state
discrimination.  As  pointed out earlier, the  generalized Bell states
are  eigenstates  of  the  unitary  operators  $X_d^{\otimes  n}$  and
$Z_d^{\dag}\otimes Z_d$,  where $d, n  \ge 2$.  We mentioned  that the
non-destructive measurement $M_1$,  effected through the ancilla $A_1$
was equivalent to measuring  an observable compatible with the unitary
operators  $X_d^{\otimes n}$,  while  the non-destructive  measurement
$M_i$  ($2 \le  i \le  n$), effected  through the  ancilla  $A_i$, was
equivalent  to measuring  an  observable compatible  with the  unitary
operators  $Z_d^{\dag}\otimes Z_d$.   That  is to  say, the  ancillary
measurements are  such that $X_d^{\otimes n}  = \exp(2\pi\iota M_1/d)$
and $Z_d^{\dag}(i-1)\otimes Z_d(i) = \exp(2\pi\iota M_i/d)$.

In  the case  of  $d=2$, of  course,  the observable  and the
unitary operator, given  by the $X\otimes  X$ and $Z\otimes Z$,  are
identical though  in general  this need  not  be the  case.  In  the
context  of distributed  computing, the  separable form  of
$X_d^{\otimes  n}$ and $Z_d^{\dag}(j)\otimes Z_d(k)$  means that
observables  compatible with them   can  be   evaluated   by  local
measurements  and   classical communication, but in so doing, the
states will of course be destroyed and  thus not  be  available
beyond the  first  measurement, so  that multiple   copies  of   the
state   would  be   necessary   for  full discrimination.   Our
circuits  overcome  this problem  by  employing quantum
communication, consisting  in the  movement of  the ancillary qubits
between  players.  Note  that  such  quantum communication  is
necessary,  since  Bell  states,  being  entangled,  possess
nonlocal correlations that cannot be accessed  locally. Further we
note that to `outsource' the  measurement of  an observable from the
system  to an ancilla, the system and ancilla  are brought into
interaction by means of a control operation (CNOT  when $d=2$) built
from the corresponding unitary  operation.  If  this  is not
entirely  clear so  far, it  is because, as is clarified  below, the
nature of this  interaction can be modified in various ways. In this
Section, we will find it convenient to  use the  notation where the
ancilla  appears to  the left  of the system qudit(s).

\begin{figure}
\includegraphics[width=3in]{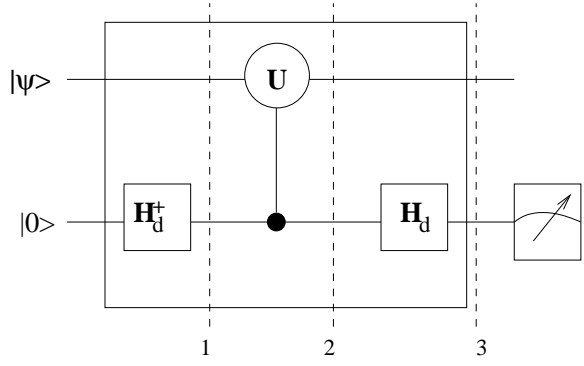}
\caption{Diagram depicting  the circuit for  `outsourcing' measurement
on a $d$  dimensional system (upper wire) to  an ancilla (lower wire).
The    circle-link   represents    the    generalized   controlled-$U$
operation. The dashed lines depict the three stages of the outsourcing
operation preceding measurement of the ancilla.}
\label{fig:utsrc}
\end{figure}

The above arguments suggest the following generalization that allow us
to go  beyond Bell state discrimination:  that for a  Hilbert space of
any finite  dimension $d \ge 2$,  an observable $W$  compatible with a
given   unitary  operator   $U$   can  be   effectively  measured   by
`outsourcing' the  measurement to  an ancilla by  means of  a suitably
generalized control-$U$ operation.  This  is the object of the Theorem
\ref{thm:utsrc}.
\begin{thm}
\label{thm:utsrc}
Given unitary operator $U$ and an observable $W$ compatible with it,
measurement of $W$ can be outsourced to an ancilla using
the controlled operation given by
${\cal C}_U \equiv \sum_j |j\rangle\langle j|
\otimes U^{j}$, where $\{|j\rangle\}$ is the possibly degenerate,
simultaneous eigenbasis of $U$ and $W$.
\end{thm}
{\bf Proof.}   The unitary operator can  in general be  written in its
diagonal basis by  $U = \sum_{j,k} e^{2\pi\iota j/d}|j;k\rangle\langle
j;k|$ ($0  \le j  \le d-1$), where  $k$ accounts for  degeneracy.  The
observable  compatible with  it is  designated to  be $W  = \sum_{j,k}
f(j)|j;k\rangle\langle  j;k|$,  where  $f(\cdot)$ is  any  real-valued
function.  The state to be measured is some $|\Psi\rangle = \sum_{k,l}
\alpha_{k,l}|k;l\rangle$   entering    the   upper   wire    in   Fig.
(\ref{fig:utsrc}). At stage 1, the state of the ancilla-system complex
is $d^{-1/2}\sum_{j,k,l}\alpha_{k,l}|j\rangle|k;l\rangle$.  Via action
of  controlled-$U$ gate,  in  stage 2,  the  state of  the complex  is
$d^{-1/2}\sum_{j,k,l}\alpha_ke^{2\pi\iota jk/d} |j\rangle|k;l\rangle$.
At  stage  3,  by the  action  of  $H_d^{\dag}$,  the above  state  is
transformed       to      $d^{-1/2}\sum_{j,k,l,m}\alpha_ke^{(2\pi\iota
j/d)(k-m)}|m\rangle|k;l\rangle               =              \sum_{k,l}
\alpha_{k;l}|k\rangle|k;l\rangle$  since  the  summation over  $j$  is
non-vanishing only when $k=m$. Therefore, a measurement on the ancilla
in  the  computational  basis   $\{|j\rangle\}$  is  equivalent  to  a
measurement of any observable $W$ on the system.  \hfill \qed

It follows  from the above that  if $|j;k\rangle$ is  an eigenstate
of $U$, then the  outsourced measurement of $W$ on  $|j;k\rangle$
will be non-destructive but return the value  $j$. This gives us the
following corollary.
\begin{crl}
\label{crl:utsrc}
If  $U_1$  and  $U_2$   are  commuting  unitary  operators,  then  the
corresponding   outsourced  observables   $W_1$  and   $W_2$   can  be
simultaneously measured.
\end{crl}
If the operator $U$ is a product of operations on subsystems, then the
control-operation can be done pair-wise on each subsystem and a common
ancilla, before  the ancilla  is finally measured.  This is  proved in
Theorem \ref{thm:decompose}.
\begin{thm}
\label{thm:decompose}
The outsourced  measurement of observable $W$  compatible with unitary
operator $U  = \bigotimes_m  U_m$, where $m$  ($=1,2,\cdots,n$) labels
the subsystems, can be performed by separate control-operations on the
individual    subsystems   $j$   from    the   same    ancilla.    The
control-operations may be performed in any order.
\end{thm}
{\bf  Proof.}  Note that  ${\cal  C}_U  =  \sum_j |j\rangle\langle  j|
\bigotimes_m (U_m)^j  = \left(\sum_j|j\rangle\langle j|\otimes (U_1)^j
\otimes                \mathbb{I}^{\otimes               (m-1)}\right)
\left(\sum_{j^{\prime}}|j^{\prime}\rangle\langle     j^{\prime}|\otimes
\mathbb{I}  \otimes   (U_2)^{j^{\prime}}  \otimes  \mathbb{I}^{\otimes
(m-2)}\right)                                                   \cdots$
$\left(\sum_{j^{\prime\prime}}|j^{\prime\prime}\rangle\langle
j^{\prime}|\otimes                                  \mathbb{I}^{\otimes
(m-1)}(U_2)^{j^{\prime\prime}}\right)$.  Therefore ${\cal C}_U = {\cal
C}_{U_1} \times  {\cal C}_{U_2}  \times \cdots {\cal  C}_{U_m}$, where
${\cal  C}_{U_k}  \equiv  \sum  |j\rangle\langle  j|\otimes  (U_k)^j$.
Since  the ${\cal  C}_{U_j}$'s commute  with each  other, they  may be
performed in any order.  \hfill \qed

However, note that though  the control operations are separable,
there is a quantum  communication of the ancilla along  the chain
formed by the players. The measurement  of $M_1$ in the preceding
Section can be   seen  as  a   special  case   of  Theorems
\ref{thm:utsrc}  and \ref{thm:decompose}.  To  see this, we  set $U
\equiv  X^{\otimes n}$, where      each     $U_i      =      X_d$.
Since      $X^{\otimes n}|\Psi_{pq_1,\cdots,q_{n-1}}\rangle      =
e^{2\pi\iota     p/d} |\Psi_{pq_1,\cdots,q_{n-1}}\rangle$,  by
Theorem  \ref{thm:utsrc}, the observable       $M_1       \equiv
\sum_{p,q_1,\cdots}       f(p) |\Psi_{pq_1,\cdots,q_{n-1}}\rangle
\langle \Psi_{pq_1,\cdots,q_{n-1}}|$  can  be  outsourced  using the
control operation ${\cal  C}_U \equiv  \sum |j\rangle\langle j|
\otimes U^j$. In view of  Eq.  (\ref{eq:genzxb}), this has the
effect: ${\cal C}_U: |j\rangle|j_1\rangle\cdots|j_n\rangle
\longmapsto |j\rangle|j_1+j\rangle\cdots|j_n+j\rangle$.   It   then
follows  from Theorem \ref{thm:decompose}  that ${\cal C}_U$ can be
broken into $n$ applications of ${\cal C}_X$  operations on an
ancilla-qudit pair, for each qudit  of the system and  a fixed
ancilla, where  ${\cal C}_X$ is precisely  the   operation  defined
in  Eq.    (\ref{eq:c-}).   In  a distributed  computing scenario,
this ancilla  must  be sequentially interacted with each system
qudit.   This clarifies our use of the Eq. (\ref{eq:c-}) as the
generalization of  the CNOT gate.  We also obtain the   general Bell
state   discrimination   circuit  described   in Eq. (\ref{aftanna})
as a  special case of Theorems \ref{thm:utsrc} and
\ref{thm:decompose}.

In general, given  any set of orthonormal states  that form a complete
basis  to an  observable  $W$, Theorem  \ref{thm:utsrc}  allows us  to
`outsource'  their measurement  to an  ancilla.   To do  so, we  first
construct a  unitary operator $U$  with respect to which  these states
are `dark',  i.e., of  which these states  are eigenstates,  and using
this  to construct  a control-$U$  operation ${\cal  C}_U$. If  $U$ is
separable,   as   is  the   case   in   our   problem,  then   Theorem
\ref{thm:decompose}  allows  ${\cal  C}_U$  to  be broken  up  into  a
sequence of pair-wise control gates.

\begin{figure}
\includegraphics[width=6in]{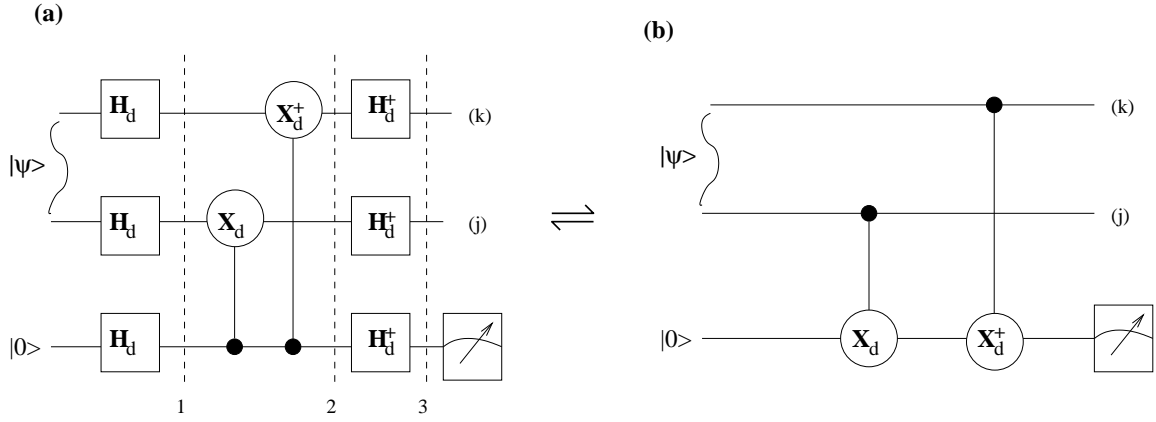}
\caption{The two circuits are equivalent to outsourcing
measurement of $Z_d \otimes Z_d^{\dag}$ on any two wires of an
$n$-qudit state. The Hadamards serve the role of reversing the
direction of control in the control gates.}
\label{fig:tvacirq}
\end{figure}

Consider    measurement   of    the    relative   parity
observable $Z_d(i-1)\otimes  Z_d^{\dag}(i)$.  Following  Theorems
\ref{thm:utsrc} and \ref{thm:decompose}, the measurement  here can
be outsourced using control-$Z_d^{\dag}$   (${\cal
C}_{Z_d^{\dag}}$)   and  control-$Z_d$ (${\cal C}_{Z_d}$) operations
from the ancilla sequentially to the two qudits.   According   to
Eq.   (\ref{eq:genzxa}),   these   require controlled-phase
operations.  However, by means of applying Hadamards, it is possible
to turn them into ${\cal C}_X$ operations. To see this, we note that
for any integer $j$,
\begin{eqnarray}
\label{eq:braek}
(Z^{\dag}\otimes Z_d)^j  &=& (Z^{\dag}_d)^j \otimes  (Z_d)^j \nonumber
\\ &=&  (H_d X^{\dag}_d  H^{\dag}_d)^j \otimes (H_d  X_d H^{\dag}_d)^j
\nonumber \\ &=& (H_d  (X^{\dag}_d)^j H^{\dag}_d) \otimes (H_d (X_d)^j
H^{\dag}_d)\nonumber\\ &=& (H_d \otimes H_d) \times (X^{\dag}_d\otimes
X_d)^j \times (H^{\dag}_d \otimes H_d^{\dag}).
\end{eqnarray}
This means that the outsourcing of measurement of $Z_d^{\dag}\otimes
Z_d$ is equivalent to the circuit in Fig. \ref{fig:tvacirq}(a),
where only ${\cal C}_X$ and ${\cal C}_X^{\dag}$ are used.
The last  result we  require says that,  by dropping the  Hadamards
in Fig.   \ref{fig:tvacirq}(a),   we    can   reverse    the control
direction. This is shown  in Theorem \ref{thm:tvacirq}. Two
advantages of such a step is  that for each outsourced measurement
of $Z^{\dag}_d \otimes Z_d$,  the number of Hadamards  is reduced by
a  factor of six and furthermore  instances of only one nonlinear
gate (namely, ${\cal C}_{X_d}$ or ${\cal C}_{X_d^{\dag}}$) need to
be used.

\begin{thm}
\label{thm:tvacirq} The two measurement circuits depicted in Fig.
\ref{fig:tvacirq} are equivalent.
\end{thm}
{\bf Proof.} Let the incoming state of the two system wires be the
pure state $|\Psi\rangle = \sum_{jk}\alpha_{jk}|j\rangle|k\rangle$ (we
ignore the fact that the summation can run on a single index on
account of Schmidt decomposability). At stage 1, the state of
the ancilla-system complex is:
$(1/\sqrt{d})\left(\sum_l|l\rangle\right)
\left(\sum_{j,k,j^{\prime},k^{\prime}}\alpha_{jk}
\exp[(2\pi\iota/d)(-jj^{\prime} + kk^{\prime})]
|j^{\prime}\rangle|k^{\prime}\rangle\right)$.
By the action of the two control-gates, the state in stage 2
is $(1/\sqrt{d})\left(\sum_{l,j,k,j^{\prime},k^{\prime}}\alpha_{jk}
\exp[(2\pi\iota/d)(-jj^{\prime} + kk^{\prime})]
|l\rangle|j^{\prime}-l\rangle|k^{\prime}-l\rangle\right)$.
In stage 3, by the action of the three Hadamards, the
state $|\Psi^{\prime}\rangle$ of the complex is
\begin{eqnarray}
\label{eq:tvacirq}
|\Psi^{\prime}\rangle &=&
(1/\sqrt{d})\left(\sum_{l,j,k,j^{\prime},k^{\prime},j^{\prime\prime},
k^{\prime\prime}}\alpha_{jk}
\exp\left[(2\pi\iota/d)(-jj^{\prime} + kk^{\prime} - ll^{\prime} +
j^{\prime\prime}[j^{\prime}-l] - k^{\prime\prime}[k^{\prime}-k])\right]
|l^{\prime}\rangle|j^{\prime\prime}\rangle|k^{\prime\prime}\rangle\right)
\nonumber \\
&=& (1/\sqrt{d})\left(\sum_{l,j,k,j^{\prime},k^{\prime},j^{\prime\prime},
k^{\prime\prime}}\alpha_{jk}
\exp[(2\pi\iota/d)(l(-l^{\prime} - j^{\prime\prime} +
k^{\prime}) + j^{\prime}[j^{\prime\prime} - j] +
k^{\prime}[k-k^{\prime\prime}])]
|l^{\prime}\rangle|j^{\prime\prime}\rangle|k^{\prime\prime}\rangle\right)
\nonumber \\
&=& (1/\sqrt{d})\left(\sum_{l,l^{\prime},j,k}\exp[(2\pi\iota/d)
(l(-l^{\prime}-j+k)]|l^{\prime}\rangle|j\rangle|k\rangle\right) \nonumber\\
&=& \sum_{j,k}\alpha_{jk}|k-j\rangle|j\rangle|k\rangle,
\end{eqnarray}
which   is  the  situation   described  by   the  circuit   in  Fig.
\ref{fig:tvacirq}(b).  In general,  the  two wires,  being  part of
a larger  system, are  in a  mixed  state. Since  a mixed  state can
be regarded  as  an  ensemble  of pure  states,  Eq.
(\ref{eq:tvacirq}) implies   the   equivalence   of   the   circuits
in   the   Fig. \ref{fig:tvacirq}(a) and  \ref{fig:tvacirq}(b) even
for  mixed states. \hfill \qed

From     Theorems     \ref{thm:utsrc},     \ref{thm:decompose} and
\ref{thm:tvacirq},  it   follows  that  the   circuitry described by
Eq. (\ref{aftannb}),  or equivalently, depicted in the second
bounded box  of  Fig. \ref{thm:tvacirq},  indeed outsources
measurement  of $Z_d\otimes Z^{\dag}_d$. More generally, Theorem
\ref{thm:tvacirq} can be  used to reverse  the direction  of control
in the  outsourcing of two-qudit observables, by replacing $U$ with
$H_d U H_d^{\dag}$ as the unitary operator on which the control gate
is based.

\section{Some applications\label{sec:pli}}
Such non-destructive state discrimination can be useful in distributed
quantum computing, especially when  there are restrictions coming from
the  topology  of the  quantum  communication  network.  Unlike  their
classical counterparts, quantum channels  are expected to be expensive
and not  amenable to change to suit  a problem at hand.  Rather, it is
worthwhile  to  use  protocols  that  minimize  quantum  communication
complexity, that is, the quantity  of quantum information that must be
communicated  between different  parties to  perform a  computation or
process some information, in a given network.

A simple  way to  perform Bell state  discrimination is for  all other
members to  communicate their qudits  to single station,  whose member
(called, say Alice) performs a  joint measurement on all $n$ qubits or
qudits to determine the state.  She then re-creates the measured state
and  transmits  them  for  further  use.   Actually,  in  the  present
situation, instead  of a  joint measurement on  all qubits,  Alice can
apply  a string  of  $n-1$ ${\cal  C}_{X}^{\dag}$  operations on  each
consecutive  pair of  qudits in  the Bell  state $|\Psi_{pq_1q_2\cdots
q_{n-1}}\rangle$ and  $H^{\dag}_d$ finally on the first  qudit.  It is
easily  seen  that  each  application of  ${\cal  C}_{X}^{\dag}$  will
disentangle the controlled  qudit from the rest. For  the Bell states,
this procedure effects the transformation:
\begin{equation}
\label{eq:transfo}
|\Psi_{pq_1q_2\cdots q_{n-1}}\rangle ~\longmapsto~
|p\rangle|q_2-q_1\rangle\cdots|q_{n-1}-q_{n-2}\rangle.
\end{equation}
Subsequent  measurement  of  each  qudit in  the  computational  basis
completely  characterizes the Bell  state. The  Bell state  thus being
discriminated, the  above procedure can  be reversed to  re-create the
state $|\Psi_{pq_1q_2\cdots  q_{n-1}}\rangle$ and transmit  it back to
the remaining players.

Irrespective  of network  topology, such  a
disentangle-and-reentangle strategy   requires   in   all   $2(n-1)$
two-qudit   gates   to   be implemented. In our  method, the number
of two-qudit  gates is the sum of  $n$  two-qudit  gates  for
determining phase  parameter  $p$  and $2(n-1)$  for  determining
the  (relative)  parities,  giving  $3n-2$ two-qudit gates.   From
this  viewpoint of consumption  of nonlinear resources, our method
does not offer any  advantage.  However, this turns  out  not to be
the  case  from  the  viewpoint  of  quantum communication
complexity.

Suppose a quantum  communication network with a star  topology and
$n$ members is given, as for  example in Fig. \ref{fig:topo}(a). For
all members to  transmit their qudits  to Alice (at  $A$), and for
her to transmit them  back would require $2(n-1)$ qudits  to be
communicated, where  the  factor 2  comes  from  the  two-way
requirement.   In  our protocol, one  way quantum  communication
suffices. For  measuring the `phase observable' $M_1$, the number of
qudits communicated is seen to be $2(n-1)$, since the ancilla must
pass through the hub to reach each member  on  a  single-edge
vertex;  and  if  measured  edgewise,  the communication  complexity
for  relative  parity  measurement  is  $n$ qudits.  In all, this
requires $3n-2$ qudits to be communicated, which is  larger  than
that  required for  a  plain  disentangle-reentangle method.

However consider a linear  configuration of the communication
network, as  in Fig. \ref{fig:topo}(b), where  members  are linked
up in  a single  series.  In the  disentangle-reentangle  method, if
Alice  is located  at  one end,  the  communication  complexity is
seen  to  be $n(n-1)$ qudits; it is $(n^2-1)/2$ if she is in the
middle.  In either case, it is of order $O(n^2)$. In contrast, our
non-destructive method can be implemented using $n-1$  qudits
communicated both for phase and relative parity measurement,
requiring  in all only $2(n-1)$ qudits to be communicated, so  that
the required communication is  only of order $O(n)$.   Thus  our
method   gives  a  quadratic  saving  in  quantum communication
complexity.

A  further  advantage, that  may  be  of  some importance  in  certain
situations, is that our method divides the required resources in terms
of  applying nonlinear  gates and  of measurements  equally  among the
various members.   In a real  life situation, this may  facilitate the
distribution  of quantum  information processing  resources  among the
various members.

\acknowledgments
We  are  thankful  to  Prof.   J.   Pasupathy,  V.  Aravindan  and  H.
Harshavardhan,   Dr.    Ashok    Vudayagiri,   Dr.    Ashoka   Biswas,
Dr. Shubhrangshu Dasgupta for useful discussions.

\appendix
\section{Closure of generalized Bell states under Hadamards\label{dx:hh}}
The action of $H\otimes H^{\dag}$ on $|\Psi_{pq}\rangle$ on the states
in Eq. (\ref{eq:cmin}) produces the effect of effectively interchanging
the indices $pq$ of $|\Psi_{pq}\rangle$:
\begin{eqnarray}
(H\otimes H^{\dag})|\Psi_{pq}\rangle &=&
\frac{1}{\sqrt{d}}\sum_{j,k,l}e^{(2\pi\iota/d)(j[p+k-l]-ql)}|k\rangle|l\rangle
\nonumber\\
&=&\frac{1}{\sqrt{d}}\sum_{j,l}e^{(2\pi\iota/d)(-ql)}|l-p\rangle|l\rangle\nonumber\\
&=&\frac{1}{\sqrt{d}}\sum_je^{(2\pi\iota/d)([d-q]l)}|j\rangle|j+p\rangle, \nonumber \\
&=& |\Psi_{q^{\prime}p}\rangle,
\end{eqnarray}
where $q^{\prime}=(d-q)~\mod~d$ and
the second step follows from noting that the only non-zero
contributions come for the case $p+k-l=0$, and an overall phase
factor has been dropped in the third step. Similarly, one finds
$(H\otimes H^{\dag})|\Psi_{pq}\rangle = |\Psi_{qp^{\prime}}\rangle$, where
$p^{\prime}=d-p$ mod $d$.

\begin{thebibliography}{10}
\bibitem{Niel}Nielsen, M.A. and Chuang, I.L.
\textit{Quantum Computation and Quantum Information,}
2000, Cambridge University Press.
\bibitem{suter}Stolze, J. and Suter, D. \textit{Quantum computing}, 2004, Wiley-Vich, ISBN
3-527-40438-4.
\bibitem{bouw}D. Bouwmeester, J.W. Pan, K. Mattle, M. Eibl, H. Weinfurter, and
A. Zeilinger, Nature 390, 575-579 (1997).
\bibitem{pan1}J.W. Pan, D. Bouwmeester, H. Weinfurter, and A. Zeilinger,
Phys. Rev. Lett. 80, 3891 (1998).
\bibitem{Mattle}K. Mattle, H. Weinfurter, P.G. Kwait, and A. Zeilinger,
Phys. Rev. Lett. 80, 3891 (1998).
\bibitem{carvalho}A.R.R. Carvalho, F. Mintert, and A. Buchleitner Phys. Rev. Lett. 93, 230501 (2004).
\bibitem{hein}M. Hein, W. Dür and H.-J. Briegel, Phys. Rev. A.
71, 032350 (2004).
\bibitem{wootters}W.K. Wootters and W.H. Zurek, Nature 299, 802 (1982).
\bibitem{walgate}J. Walgate, A.J. Short, L. Hardy and V. Vedral,
Phys. Rev. Lett. 85, 4972 (2000).
\bibitem{gosh1} S.Ghosh, G. Kar, A.Roy, A.S. Sen(De) and U. Sen,
Phys. Rev. A. 87, 277902 (2001).
\bibitem{vermani}S. Virmani, M.F. Sacchi, M.B. Plenio and D. Markham,
Phys. Lett. A 288, 62(2001).
\bibitem{chen}Y.X. Chen and D. Yang, Phys. Rev. A 64, 064303 (2001).
\bibitem{gosh2}S.Ghosh, G. Kar, A.Roy and D. Sarkar, Phys. Rev. A. 70, 022304 (2004).
% \bibitem{lutk}N. Lutkenhaus, J. Calsamiglia and K.A. Suominen,
% Phys. Rev. A 59, 3295 (1998).
\bibitem{cola}M.M. Cola and M.G.A. Paris, Phys. Lett. A 337, 10 (2005).
\bibitem{pan2}J.W. Pan and A. Zeilinger, Phys. Rev. A 57, 2208 (1998).
\bibitem{kim}Y.H. Kim, S.P. Kulik and Y. Shih, Phys. Rev. Lett. 86, 1370 (2001).
\bibitem{boschi}D. Boschi, S. Branca, F.D. Martini, L. Hardy and S. Popescu,
Phys. Rev. Lett. 80, 1121 (1998).
\bibitem{kni96} E. Knill, eprint quant-ph/9608048.
\bibitem{ben93} C. H. Bennett, G. Brassard G, C. Crépeau C, R. Josza R, A.
Peres and W. K. Wootters Phys. Rev. Lett. 70, 1895 (1993).
%\bibitem{ral02} T. C. Ralph, N. K. Langford, T. B. Bell, and A. G. White,
%Phys. Rev. {\bf A 65}, 062324.
\end{thebibliography}
\end{document}